\documentclass[aps,twocolumn,showpacs,amssymb,pre]{revtex4-1}
\usepackage[dvips]{graphicx}
\usepackage{dcolumn}
\usepackage{bm}
\usepackage[dvips]{epsfig}
\usepackage{color}
\usepackage{ulem}

\newcommand \be {\begin{equation}}
\newcommand \ee {\end{equation}}
\newcommand \bea {\begin{eqnarray}}
\newcommand \eea {\end{eqnarray}}

\begin{document}
\title{Granular flow through an aperture: influence of the packing fraction}

\author{M. A. Aguirre$^{1}$, R. De Schant$^{1}$, and J.-C. G\'{e}minard$^{2}$}
\affiliation{$^{1}$Grupo de Medios Porosos, Fac. de Ingenier\'{\i}a, Universidad de Buenos Aires.
Paseo Col\'{o}n 850, (C1063ACV) Buenos Aires, Argentina.\\
$^{2}$Universit\'e de Lyon, Laboratoire de Physique, Ecole Normale
Sup\'erieure de Lyon, CNRS, 46 All\'ee d'Italie, 69364 Lyon cedex
07, France.}

\begin{abstract}

For the last 50 years, the flow of a granular material through an
aperture has been intensely studied in gravity-driven vertical
systems (e.g. silos and hoppers). Nevertheless, in many industrial
applications, grains are horizontally transported at constant
velocity, lying  on conveyor belts or floating on the surface of
flowing liquids. Unlike fluid flows, that are controlled by the
pressure, granular flow is not sensitive to the local pressure but
rather to the local velocity of the grains at the outlet. We can
also expect the flow rate to depend on the local density of the
grains. Indeed, vertical systems are packed in dense configurations
by gravity but, in contrast, in horizontal systems the density can
take a large range of values, potentially very small, which may
significantly alter the flow rate. In the present article, we study,
for different initial packing fractions, the discharge through an
orifice of monodisperse grains driven at constant velocity by a
horizontal conveyor belt. We report how, during the discharge, the
packing fraction is modified by the presence of the outlet and we
analyze how changes in the packing fraction induce variations in the
flow rate. We observe that variations of packing fraction do not
affect the velocity of the grains at the outlet and, therefore, we
establish that flow-rate variations are directly related to changes
in the packing fraction.

\end{abstract}
\maketitle

\section{Introduction and background}

Because of its obvious practical relevance,
the flow of granular media through an aperture has been intensely studied
in the last $50$ years in vertical gravity-driven systems (e.g. silos and hoppers)
\cite{Beverloo,Kadanoff,deGennes,Trappe,Jaeger,Duran,Ristow,Nedderman,Tuzun,Savage,Tighe}.
The discharge of a silo through an orifice can present three
regimes: a continuous flow, an intermittent flow, or a complete
blockage due to arching \cite{Mankoc07,Mankoc2009,Ulissi}.

In the continuous flow regime, the \textit{mass flow rate} $Q_\mathrm{m} \equiv dM/dt$
(i.e. the mass $M$ flowing out per unit time~$t$)
is generally satisfactorily given by the so-called Beverloo's law \cite{Beverloo,BrownBook}:
$Q_\mathrm{m} = C \rho_\mathrm{3D} \sqrt{g}(A-k\,D)^{5/2}$
where $A$ is the diameter of the opening (assumed here to be circular),
$\rho_\mathrm{3D}$ the bulk density
of the granular sample, $g$~the acceleration due to gravity and $D$
the diameter of the granules whereas $k$ and $C$ are empirical,
dimensionless, constants.
The Beverloo's law thus points out a value $A_\mathrm{c} \equiv k\,D$
of the aperture size $A$ at which the flow rate is expected to vanish.
Therefore, instead of $A$, the effective aperture $A_\mathrm{eff} \equiv A - k\,D$ is to be considered.
The value of~$k$ has been found to be independent of the size $D$ of the grains
and to take values ranging from 1 to 3 depending on the grains and container properties \cite{Nedderman1980}.
Nevertheless, some works \cite{Zhang, Mankoc07} claim that the only plausible value for $k$ is~$1$.
It should also be noted that a recent work \cite{Sheldon} states that $k$ is just a fitting parameter
with no clear physical meaning as the authors found clogging of the flow for apertures $A > k\,D$.
In the jamming regime, the jamming probability has been shown to be
controlled by the ratio $A/D$ of the aperture size to the grain
diameter \cite{Mankoc2009,To,Zuriguel03,Zuriguel05,Janda,Ulissi}.

In many industrial applications, however, granular materials are
transported horizontally, lying on conveyor belts \cite{desong03} or
floating on the surface of flowing liquids
\cite{Guariguata2009,Guariguata2012,Lafond2013}. In a
two-dimensional (2D) configuration -- or similarly for slit shaped
apertures -- one expects Beverloo's law to be: $Q_\mathrm{m} = C
\rho_\mathrm{2D} \sqrt{g}\,(A - k\,D)^{3/2}$ \cite{BrownBook}.
Recent works considered the discharge of a dense packing of disks
driven through an aperture by a conveyor belt. For large apertures
($A/D \ge 6$), the flow-rate is continuous throughout the discharge.
In this case, the number of discharged disks  $N$ depends linearly
on time~$t$ and the flow rate $Q \equiv dN/dt$ (i.e. the number $N$
of disks flowing out per unit time $t$) obeys:
\begin{equation}
Q = C\,\Bigl(\frac{4}{\pi D^2}\Bigr)\,V\,\Bigl( A - k\,D \Bigr)
\label{beverloo}
\end{equation}
where $k \simeq 2$ and the constant $C$ reduces to the packing
fraction \cite{Aguirre2010}. Indeed, ${\pi D^2}/{4}$ is the surface
area of one disk so that $C({4}/{\pi D^2})$ is the number of grains
per unit surface which, multiplied by the belt velocity and by the
size of the aperture, gives an estimate of the number of disks
flowing out per unit time. Note that Eq.~(\ref{beverloo}) is
equivalent to the 2D Beverloo's law in which the typical velocity
$\sqrt{g\,A_\mathrm{eff}}$, understood as the typical velocity of
the grains at the outlet, is replaced by the belt velocity $V$. It
predicts that the dimensionless flow rate $Q^*\equiv Q D/V$ is
independent of $V$ and increases linearly with the dimensionless
aperture-size $A/D$. It is interesting to note that this empirical
law was demonstrated to be valid for small apertures $A/D < 6$, even
if the system is likely to jam and deviations from linearity might
be expected \cite{Aguirre2010}. Indeed, in 3D configurations, a
marked deviation from the $5/2$ Beverloo's scaling has been observed
for very small apertures \cite{Mankoc07}. Moreover, these previous
works show that, unlike fluid flows, granular flows are not governed
by the pressure, but rather controlled by the velocity of the grains
at the outlet \cite{Aguirre2010, Aguirre2011}. The latter does not
necessarily depend on the stress conditions in the outlet region as
proven by the experimental fact that, in gravity-driven systems, the
typical velocity at the outlet is $\sqrt{g\,A}$, independent of the
pressure. These observations were corroborated in vertical
gravity-driven systems \cite{Perge2012}.

Even if the Beverloo's law has been intensively discussed, the
influence of the packing fraction, i.e. the ratio of the area
occupied by grains over the total available area, has only been
partially considered. However, it is expected that the flow rate can
be altered by the packing fraction of the grains aside from their
velocity. On the one hand, vertical granular systems are usually
gravity packed in dense configurations, except in situations where
inflow rate is controlled \cite{Huang2006,Huang2011}, and little
effect of the packing fraction is expected in usual conditions. But,
on the other hand, in horizontal configurations the packing fraction
can explore a large range of values and one can expect significant
changes in the flow rate. Ahn {\it et al} studied granular flow rate
in vertical silos filled under different conditions, which, as a
consequence, lead to different values of packing
fraction~\cite{Ahn2008}. However, aiming at relating flow-rate
variations to changes in the pressure, they do not discuss the
possibility that the variations could be due to changes in the
packing fraction itself. In a more recent work, Janda {\it et al}
studied velocity and packing fraction profiles at the outlet and
they obtained a new expression, independent of $k$, for the granular
flow rate \cite{Janda2012}.

In the present article, we study the discharge of monodisperse acrylic rings,
driven through an orifice, at a constant velocity, by a horizontal conveyor belt.
For various initial packing fractions, we report simultaneous measurements of the
grains velocity, packing fraction and flow-rate throughout the discharge process.

\section{Setup and protocol}

\begin{figure}[ht]
\includegraphics [width=\columnwidth]{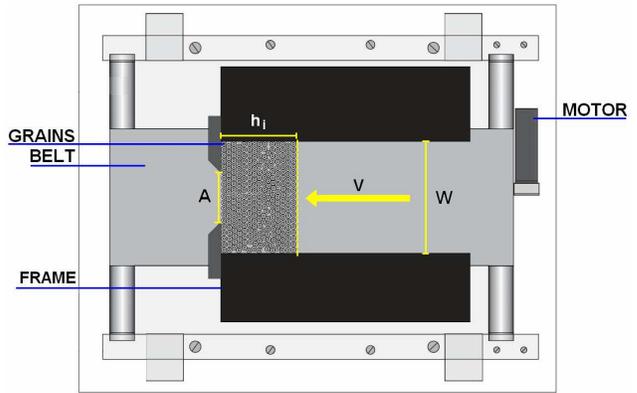}
\caption {Sketch of the experimental setup.} \label{ExpSetup}
\end{figure}

The experimental setup (Fig.~\ref{ExpSetup}) consists of a conveyor
belt made of black paper (width $11$~cm, length $34.5$~cm) above
which a confining cardboard frame (inner width $9$~cm, length
$20$~cm) is maintained at a fixed position in the frame of the
laboratory. A motor drives the belt at a constant velocity $V$. The
granular material is made of {acrylic} rings of thickness $e = (2.00 \pm 0.01)$~mm
and external diameter $D = (4.00 \pm 0.01)$~mm.

Downstream, the confining frame exhibits, at the center, a
sharp aperture of width $A$.
The aperture width can be tuned up to $9$~cm but we shall report data obtained
for a {single} width $A = (4.1 \pm 0.1)$~cm.
The aperture size $A$ is of about 10
times the grain diameter $D$, so that the condition insuring the
continuous flow, $A/D \ge 6$, is satisfied~\cite{Aguirre2010}.

The grains are imaged from top by means of a digital scanner (Canon, CanoScan LIDE200)
placed upside down above the frame.
In order to focus on the top of the grains without mechanical contact (gap of about 1~mm)
and thus avoid friction between the grains and the scanner window, the latter has been replaced by a thinner one.
The use of a scanner has the advantage of avoiding optical
aberrations {and makes it possible to} obtain, for cheap, homogeneously lighted images with a high
resolution ($12$~pixels/mm, the grain diameter being thus of the order of 50 pixels).

Before the flow is started, the initial state of the system is
obtained by placing inside the confining frame, in a disordered
manner, $N_0 = 350$  grains which initially cover the surface area
$S = W h_i$, where $W$ is the inner width of the frame ($W = 9$~cm)
and, thus, $h_i$ the length in the flow direction that is initially
covered with grains. We prepare systems with different initial
packing fractions: ~$<C_i> = 0.81 \pm 0.02$~($h_i \simeq 6$~cm);
~$<C_i> = 0.66 \pm 0.02$ ~($h_i \simeq 7.5$~cm); ~$<C_i> = 0.46 \pm
0.03$ ~($h_i \simeq 10.5$~cm) and ~$<C_i> = 0.38 \pm 0.06$ ~($h_i
\simeq 13.0$~cm). The homogeneity of the initial packing throughout
the system is controlled by measuring {the packing fraction} {along
the flow direction} in successive layers of {width $W$ and thickness
$2\,D$}. Grains are locally rearranged if {the packing fraction is}
not within $10\%$ of the {chosen} average $<C_i>$.

The discharge is then initiated by setting the belt velocity $V$ to
a chosen value. Experiments {were} performed using six different
values of $V$:~$(3.6 \pm 0.2)$~mm/s~[$0.9\,D\,s^{-1}$];~$(8.7 \pm
0.3)$~mm/s ~[$2.2 D\,$s$^{-1}$];~$(9.6 \pm
0.2)$~mm/s~($2.4\,D\,$s$^{-1}$);~$(11.3 \pm
0.3)$~mm/s~[$2.8\,D\,$s$^{-1}$] and~$(13.4 \pm
0.6)$~mm/s~[$3.3\,D\,$s$^{-1}$].
The evolution of the discharge process is assessed by repetitively
moving the belt at the chosen constant velocity $V$ during a time interval $dt=0.1$~s and
by recording an image from the scanner while the belt is at rest.

For the present study the image analysis is used to determine {the}
packing fraction{, $C$,} and the number of grains, $N_\mathrm{in}$,
that remain inside the confining frame at time $t$. To do so, an
intensity threshold is used to convert each image into binary: white
is assigned to the rings (grains) and black is assigned to the
background. Therefore, black disks at the center of each grains are
isolated from one another, which makes it easy to detect them and to
compute the number of grains remaining in the frame,
$N_\mathrm{in}$, or, equivalently, the number of disks that flowed
out the system at time $t$, $N \equiv N_0 - N_\mathrm{in}$. The
instantaneous flow-rate (averaged over $dt=0.1$~s, because of the
acquisition rate) is defined as $Q = dN/dt$.

The packing fraction $C$ is, by definition, the fraction of the surface area covered by the grains.
In order to measure $C$, the black disk at the center of the rings is filled with white in order to obtain white disks.
The number of white pixels over the total number of pixels in the region of interest is a direct measurement of $C$.

The reproducibility of the experiments has been checked by repeating
the procedure up to three times for each set of the control
parameters ($<C_i>$, $V$).

\section{Experimental results}

\subsection{Flow rate}
\label{Q}

\begin{figure}[ht]
\includegraphics[width=\columnwidth]{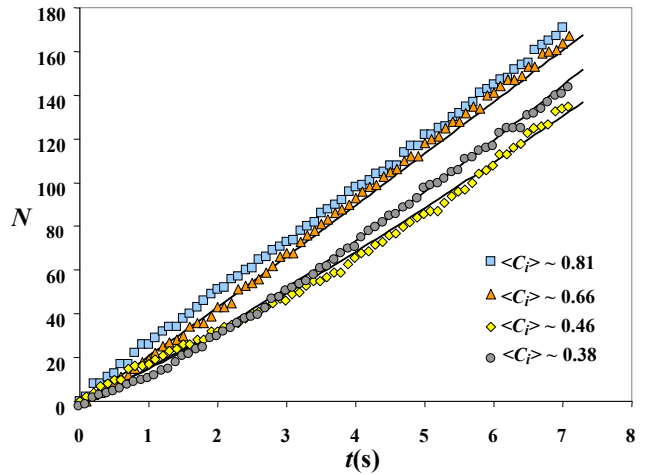}
\caption {Number of grains $N(t)$ vs. time $t$ for $V = 9.6$~mm/s
and different initial packing fractions $C_i$. The number $N(t)$ is
linear in $t$ for {initially-dense systems} ($<C_i> \sim 0.81$)
indicating a constant flow-rate. For initially loose systems, $N(t)$
exhibits a non linear dependence on time $t$, which is explained by
the increase of the packing fraction in the outlet region. Solid
lines correspond to fitting curves obtained with Eq.~(\ref{Nt}).
$<C_i> \sim 0.66 $ is fitted with $C_i=0.65 \pm 0.01$, $\alpha =
(0.90 \pm 0.06)~$s$^{-1}$ leading to $\lambda = (1.1 \pm 0.1)$~cm
and $\beta/V=(31 \pm 1)$~cm$^{-1}$. $<C_i> \sim 0.46$ is fitted with
$C_i=0.49 \pm 0.01$, $\alpha = (0.34 \pm 0.1)~$s$^{-1}$ leading to
$\lambda = (2.8 \pm 0.2)$~cm and $\beta/V=(30 \pm 1)$~cm$^{-1}$.
$<C_i> \sim 0.38$ is fitted with $C_i=0.40 \pm 0.01$, $\alpha =
(0.31 \pm 0.04)~$s$^{-1}$ leading to $\lambda = (3.5 \pm 0.5)$~cm
and $\beta/V=(31 \pm 1)$~cm$^{-1}$.} \label{No-t-V9mm}
\end{figure}

The discharge process is analyzed as long as grains fill a distance
of $2D$ upstream of the outlet. We report the number of grains that
flowed out the system, $N$ as a function of time $t$. Two types of
behavior are observed (Fig.~\ref{No-t-V9mm}):

\begin{enumerate}
\item For initially dense systems, $N$ increases linearly with the time $t$. The flow-rate $Q$ is constant \label{II}
\item For initially loose systems, $N$ does not increase linearly with time $t$. The flow-rate $Q$ varies during the discharge. \label{I}
\end{enumerate}

\begin{figure}[ht]
\includegraphics [width=\columnwidth]{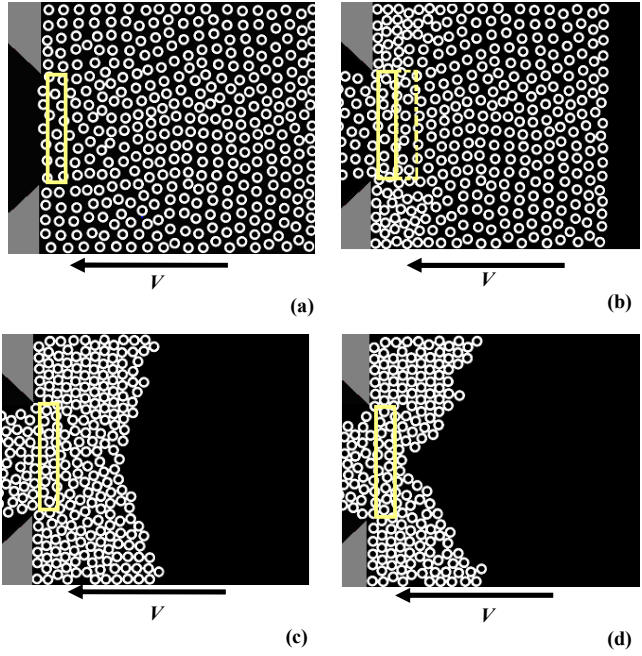}
\caption {Snapshots of the system during the discharge process for a
system with ~$<C_i> = 0.46 \pm 0.03$ driven at $V=(8.7 \pm
0.3)$~mm/s. The arrows indicate the flow direction. (a) and (b)
correspond to $t=0$~s and $t=2$~s, the first stage of the process
(transient stage): the grains are piling progressively and the flow
rate depends on time. (c) and (d) correspond to $t=9.3$~s and
$t=12.8$~s, the second stage of the process: the system has reached
a steady packing fraction and the flow rate $Q$ remains constant.
The solid box in each image encloses the region of surface area
$2\,D\,A$ upstream of the outlet (of size $A$) in which the packing
fraction is measured. Note that (d) corresponds to the last image
considered for the analysis. The dashed box in (b) indicates the
upstream region in which we define $C_\mathrm{vic.}$ in the model.}
\label{discharge}
\end{figure}

\noindent The difference can be easily understood by considering
that, for initially loose systems, the grains are progressively
piling against the downstream wall Fig.~\ref{discharge}. The
discharge process can be thus described in two stages:
\begin{itemize}
\item First stage (\textbf{transient}): the grains are piling progressively and the flow rate $Q$ depends on time.
\item Second stage (\textbf{steady}): the system has reached a steady packing fraction, $C_\infty$,
slightly smaller than the maximum possible packing fraction
$C_\mathrm{max} = 0.82$ (corresponding to the close packing), and
the flow rate $Q$ remains constant.
\end{itemize}

\subsection{Packing fraction}
\label{sec:C}

We expect the flow rate to be influenced by the packing
fraction near the outlet. Therefore, we measure the packing fraction
upstream of the aperture, in a region of width $A$ {and thickness $2\,D$}.
The region under analysis is highlighted by a solid box in each of the images in Fig.~\ref{discharge}.

\begin{figure}[ht]
\includegraphics[width=\columnwidth]{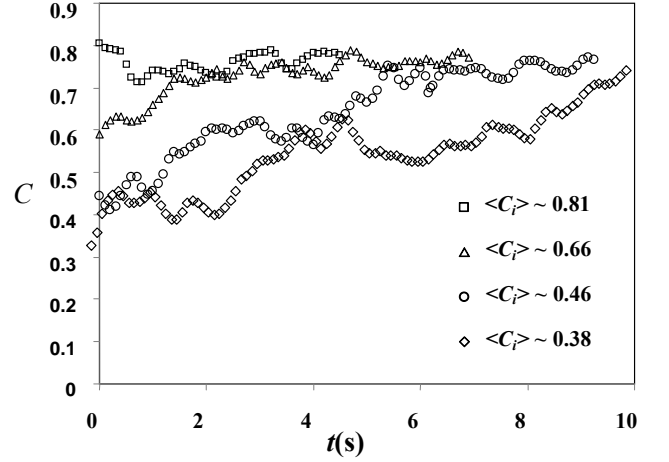}
\caption {Packing fraction $C$ in the outlet region vs. time $t$. We
observe that the temporal evolution of $C$ depends strongly on
its initial value $C_i$ ($V = 9.6$~mm/s).}
\label{CvsCi}
\end{figure}

\begin{figure}[ht]
\includegraphics[width=\columnwidth]{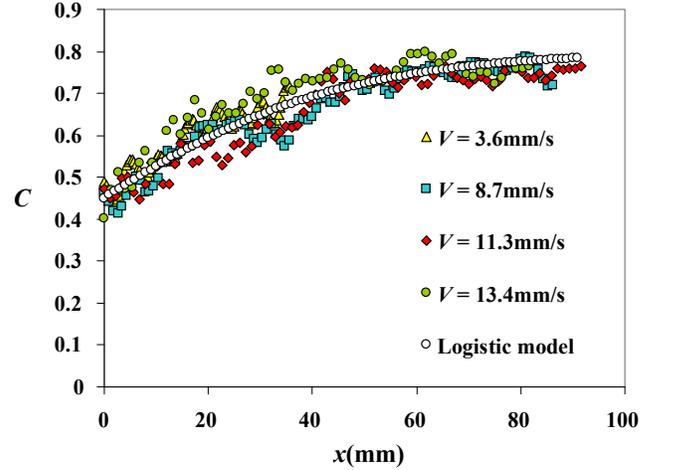}
\caption {Packing fraction $C$ in the outlet region as a function of
the distance traveled by the belt $x = V\,t$. A nice collapse of the
experimental results is observed. The dotted line corresponds to the
logistic model, Eq.~(\ref{Cx}) with: $C_\infty =0.8$, $C_i = 0.45$
and $\lambda = (2.5 \pm 0.5)$~cm.} \label{CvsV-Ci0p46}
\end{figure}

During the discharge process, the grains pile progressively against
the downstream wall until a steady state is reached. Accordingly, we
observe that the packing fraction $C$ increases up to the asymptotic
value, $C_\infty \sim 0.8$, slightly smaller than the value
$C_\mathrm{max} = 0.82$ corresponding to the close packing
(Figs.~\ref{CvsCi}~and~\ref{CvsV-Ci0p46}).

We observe that the temporal evolution of the packing fraction
strongly depends on the initial {packing fraction}
(Fig.~\ref{CvsCi}) and, as expected, the asymptotic value is reached
faster for larger belt velocities, $V$. Indeed, for a given initial
$<C_i>$, all curves collapse when $C$ is reported against $x =
V\,t$, the distance traveled by the belt at time $t$
(Fig.~\ref{CvsV-Ci0p46}).

\section{Discussion and Conclusions}
\label{Discussion}

We aim here at accounting for the temporal evolution of the packing
fraction in region close to the outlet, $C(t)$.

On the one hand, it is expected that the packing fraction increase,
due to grains that enter the outlet region from the upstream region,
at a rate $r_\mathrm{in}$ which should be proportional to:
\begin{itemize}
\item the belt velocity $V$: the higher the value of $V$ the larger the income of grains from the upstream region;
\item the packing fraction in the vicinity $C_\mathrm{vic.}$ upstream of the outlet,
i.e the region enclosed in the dashed box in
Fig.~\ref{discharge}(b): a larger packing fraction indicates a
larger amount of grains accessing from the upstream region;
\item  the available space, thus to the difference between the $C$
and its maximum accessible value, $C_\mathrm{max}$: more available
space allows a larger income of grains from the upstream region.
\end{itemize}

In addition, as can be observed in Fig.~\ref{discharge} (b), we can
further assume that the packing fraction in the vicinity of the
outlet does not differ significantly from that in the outlet region
and we take $C_\mathrm{vic.} \simeq C$. We thus write:
\begin{equation}
r_\mathrm{in} = \beta_\mathrm{in}\,V\,(C_\mathrm{max} - C)\,C.
\label{rin}
\end{equation}

On the other hand, $C$ is expected to decrease, due to the grains
that flow out through the aperture, at a rate $r_\mathrm{out}$
proportional to:
\begin{itemize}
\item $V$: the higher the value of $V$ the larger the outflow from
the system;
\item the local packing fraction $C$: a larger packing fraction at the
outlet indicates a larger amount of grains leaving the system.
\end{itemize}

Therefore:
\begin{equation}
r_\mathrm{out} = -\beta_\mathrm{out}\,V\,C.
\label{rout}
\end{equation}

Collecting Eqs.~(\ref{rin}) and~(\ref{rout}), we obtain the net variation of the packing fraction in the form :\begin{equation}
\frac{dC}{dt}= \alpha\,C\,( 1 - C / C_\infty )
\label{equa_diff}
\end{equation}
where $C_\infty = C_\mathrm{max} - \beta_\mathrm{out}/\beta_\mathrm{in}$ and $\alpha = \beta_\mathrm{in}\,C_\infty\,V$.

Taking into account the initial condition that $C(0) = C_i$,
the solution {of Eq.~(\ref{equa_diff}) can be written in the form}:
\begin{equation}
C(t) = \frac{C_\mathrm{\infty}}{1 +
\frac{C_\mathrm{\infty}-C_i}{C_i}\,e^{- \alpha\,t}}.
\label{C}
\end{equation}

We {point out} that the prefactor $\alpha$
is proportional to the belt velocity $V$ which
provides the only timescale of the problem. This assertion is
compatible with the observation of a nice collapse of the
experimental data observed when the packing fraction in the outlet region
is reported as a function of the distance traveled by the belt $x = V\,t$ (Fig.~\ref{CvsV-Ci0p46}).
Therefore Eq.~(\ref{C}) can be rewritten as:
\begin{equation}
C(x) = \frac{C_\mathrm{\infty}}{1 +
\frac{C_\mathrm{\infty}-C_i}{C_i}\,e^{-x/\lambda}} \label{Cx}
\label{CvsX}
\end{equation}
with $\lambda$ a characteristic travel distance which is thus
independent of the velocity. The measurements of $C$
(Fig.~\ref{CvsV-Ci0p46}) are satisfactorily described by
Eq.~(\ref{Cx}). For instance, the interpolation of the experimental
data for all velocity $V$ leads to $\lambda = (2.5 \pm 0.5)$~cm
($\sim\,0.6\,A$) and $C_\mathrm{\infty} = 0.80$ for $C_i = 0.45$. We
indeed observe that the steady value of the packing fraction
$C_\mathrm{\infty}$ is smaller than $C_\mathrm{max}$ as expected
from our simple description of the problem.

Later, we will discuss the meaning of this characteristic length
$\lambda$ and its dependence with the initial packing fraction. But
now, it is particularly interesting to analyze the potential effects
of the changes in the local packing fraction $C$ on the flow rate.
To do so, we consider that the flow rate $Q$ is proportional to $C$
and $V$.

\begin{figure}[ht]
\includegraphics[width=\columnwidth]{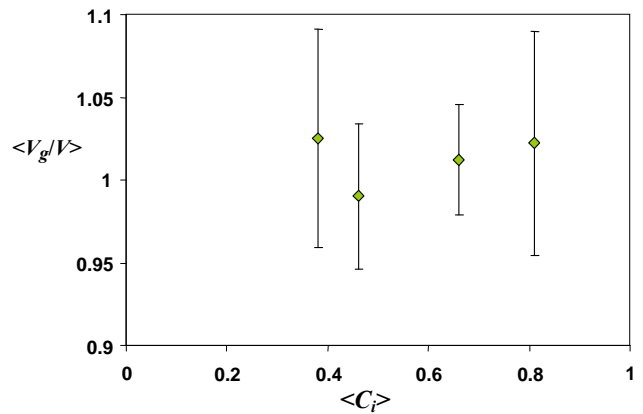}
\caption {Velocity $V_\mathrm{g}$ of the grains upstream of the
outlet vs. $C_i$. The velocity is averaged over the duration of the
discharge and normalized with the belt velocity. Even if the packing
fraction increases during the discharge, i.e. for systems with $C_i
< 0.8$, grain velocities oscillate within $7\%$ of the velocity of
the conveyor belt.} \label{V vs Ci}
\end{figure}

We report in Fig.~\ref{V vs Ci} the average velocity,
$V_\mathrm{g}$, of the grains in the region upstream the outlet
(Fig.~\ref{discharge}). We display the average over the duration of
the discharge. We observe that $V_\mathrm{g}$ almost equals the belt
velocity (to within the experimental uncertainty). No systematic
dependence is observed as a function of $C_i$, which indicates that
this average is not altered by the presence or the absence of a
transient. Therefore we can state that the characteristic velocity
of the grains at the outlet remains approximately constant and equal
to the belt velocity during the entire discharge. Moreover, we have
observed that the instantaneous velocity, even if the measurements
are noisier, does not significantly deviate from $V$. Thus, the
variations of the flow rate can only be attributed to the changes in
the local packing fraction $C$.

With the above statement in mind, we can replace the constant
packing fraction in Eq.~(\ref{beverloo}) by the time-dependent
packing fraction given by Eq.~(\ref{C}). Doing so, we get the number
of grains that left the system at time $t$ in the form:
\begin{equation}
N(t)=  C_{\infty} \beta \Bigl\{t - \frac{1}{\alpha}\,\ln\Bigl[\frac{C(t)}{C_i}\Bigr]\Bigr\}
\label{Nt}
\end{equation}
with $\beta=\frac{4 V}{\pi D^2}(A-k\,D)$. A good agreement of
experimental data with Eq.~(\ref{Nt}) (solid lines in
Fig.~\ref{No-t-V9mm}) is observed. We found that $<k>=0.8 \pm 0.4$
($<\frac{\beta}{V}>= (30 \pm 2)$~cm$^{-1}$) and, as will be
explained below, we also observed that values of $\lambda=V/\alpha$
depend on $C_i$ . The agreement confirms that the typical velocity
of the grains at the outlet to be considered in Berverloo's law is
not altered by the local packing fraction $C$. It should also be
noted that for initially dense systems the second term in
Eq.~(\ref{Nt}) vanishes and Beverloo's law (Eq.(\ref{beverloo}))
with a constant $C=C_{\infty}$ is retrieved:
$Q=\frac{dN}{dt}=C_{\infty} \beta $. Actually, in this case, $\alpha
\rightarrow \infty $ and $\lambda \rightarrow 0$ and a linear
regression corroborates that $<\frac{\beta}{V}>= (30 \pm
1)$~cm$^{-1}$.

\begin{figure}[ht]
\includegraphics[width=\columnwidth]{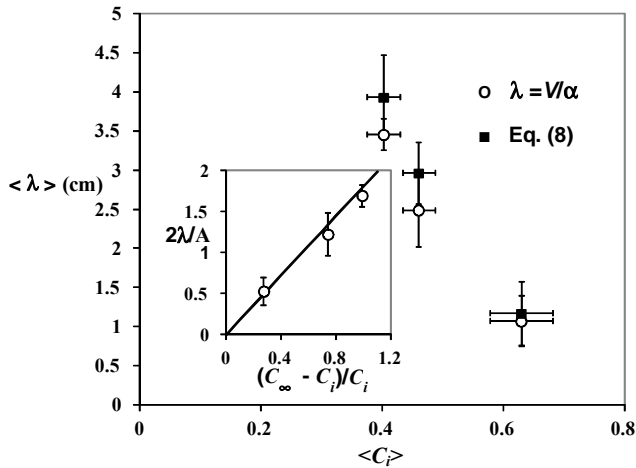}
\caption {$<\lambda>$ as a function of $<C_i>$ is presented.
($\circ$) corresponds to values of $\lambda=\frac{V}{\alpha}$
obtained with $\alpha$ values from fitting experimental data with
Eq.~(\ref{Nt})  and ($\blacksquare$) corresponds to $\lambda$ values
obtained from Eq.~(\ref{LvsCi}). Values of $<C_i>$ are mean values
obtained from fitting experimental data with Eq.~(\ref{Nt}). Inset:
experimental values $\lambda=\frac{V}{\alpha}$ are fitted with
Eq.~(\ref{LvsCi}), slope is found to be $1.7 \pm 0.1$ in accordance
with the expected value $\frac{W}{W-A}=1.8\pm 0.1$(solid line), it can be
observed that for $C_i=C_\mathrm{\infty}$ effectively $\lambda =
0$.} \label{LvsCiFig}
\end{figure}

As for the meaning of the characteristic length $\lambda$, it
corresponds to the travel distance over which the system reaches the
steady state (Eq.~\ref{CvsX}). It can be estimated by considering
that the packing fraction, in a region above the downstream wall of
typical height $A/2$ (which corresponds to the typical height of the
arch that forms above the outlet), must have reached its
steady-state value (of about $C_\mathrm{max}$) for $x \sim \lambda$.
In order to get a crude estimate, neglecting the outflow, one can
assume that a region of height $A/2 + \lambda$ and packing fraction
$C_i$ is compacted in a region of height $A/2$ and packing fraction
$C_\mathrm{max}$, which leads to $\lambda \sim
(A/2)\,(C_\mathrm{max}-C_i)/C_i$. This estimate is compatible with
the increase of $\lambda$ when $C_i$ is decreased (see
Fig.~\ref{CvsCi}) and with the absence of significant transitory for
$C_i \sim C_\mathrm{max}$. In our experimental configuration, the
outflow cannot be neglected as the width of the system $W$ is not
much larger than the aperture size $A$ and the maximum packing
fraction that can be reached is $C_\mathrm{\infty}\cong 0.8$. In
order to take into account the grains that escape the system, one
can add a correction  factor and write:
\begin{equation}
 \lambda \approxeq (A/2)\,(C_\mathrm{\infty}-C_i)/[C_i(1-A/W)]
\label{LvsCi}
\end{equation}
with parameters $\alpha$ and $C_i$ obtained from fitting the
experimental data with Eq.~(\ref{Nt}), mean values of
$\lambda=\frac{V}{\alpha}$ as a function of $<C_i>$ are shown in
Fig. \ref{LvsCiFig} and are in agreement with values of $\lambda$
obtained with Eq.~(\ref{LvsCi}).

In summary, we have simultaneously measured the flow rate and the
packing fraction in the outlet region of a discharging 2D-silo. We
have observed that, for initially loose systems, the packing
fraction {in the outlet region} evolves during the discharge and
that, at the same time, the flow rate is not constant. We proposed
that the flow rate is directly altered by the variations of the
local density of the granular material and not by variations of the
typical velocity at the outlet. This assertion is supported by a,
simplistic, logistic model, accounting for the temporal evolution of
both the packing fraction and the flow-rate, which proved to be in
agreement with our experimental data.

\acknowledgments This work has been supported by the program UBACyT
$2002010010089$ (UBA) and the International Cooperation Program
CONICET-CNRS. M. A. A. acknowledge support from CONICET.

\section*{Appendix: derivation of Eq.~(7)}
\appendix*

As explained in Sec.~\ref{Discussion}, we can replace the constant
packing fraction in Eq.~(\ref{beverloo}) by the time-dependent
packing fraction given by Eq.~(\ref{C}):
\begin{equation}
Q \equiv dN/dt = C(t)\,\Bigl(\frac{4}{\pi D^2}\Bigr)\,V\,\Bigl( A -
k\,D \Bigr) \label{beverloo2}
\end{equation}
Therefore $N(t)$ can be obtained by integrating the above
expression between $0$ and $t$:
\begin{equation}
N(t) = N(0)+\,\Bigl(\frac{4}{\pi D^2}\Bigr)\,V\,\Bigl( A - k\,D \Bigr)
\int\frac{C_\mathrm{\infty}}{1 +
\frac{C_\mathrm{\infty}-C_i}{C_i}\,e^{- \alpha\,t}}dt
\label{C2}
\end{equation}
The following substitution can be made $y=A e^{- \alpha\,t}$ with $A=\frac{C_\mathrm{\infty}-C_i}{C_i}$ leading to:
\begin{equation}
\int C(t) dt=
\frac{1}{\alpha}ln\Bigl( \frac{1+A e^{- \alpha\,t}}{A e^{- \alpha\,t}} \Bigr)
\end{equation}
which evaluated between $0$ and $t$ is:
\begin{equation}
\int C(t) dt=
\frac{1}{\alpha}ln\Bigl[ \Bigl(  \frac{1+A e^{- \alpha\,t}}{A e^{- \alpha\,t}}\Bigr)\Bigl(\frac{A}{1+A} \Bigr)\Bigr]
\end{equation}
Regarding that $1+A=\frac{C_\mathrm{\infty}}{C_i}$ and $1+A e^{- \alpha\,t}=\frac{C_\mathrm{\infty}}{C(t)}$:
\begin{equation}
\int C(t) dt= t-
\frac{1}{\alpha}ln \Bigl(\frac{C(t)}{C_i} \Bigr)
\end{equation}
So, we finally arrive to Eq.~(7) by considering $N(0)=0$, i.e. there are no disks flowing out of the system at $t=0$:
\begin{equation}
N(t)=  C_{\infty} \beta \Bigl\{t -
\frac{1}{\alpha}\,\ln\Bigl[\frac{C(t)}{C_i}\Bigr]\Bigr\}
\label{Nt-Ap}
\end{equation}

\end{document}